\documentclass[iop,revtex4]{emulateapj}
\usepackage[dvips]{color}
\begin{document}

\title{SOFIA/EXES Observations of Water Absorption in the Protostar AFGL 2591 at High Spectral Resolution}

\author{Nick Indriolo\altaffilmark{1,2},
D. A. Neufeld\altaffilmark{1},
C. N. DeWitt\altaffilmark{3},
M. J. Richter\altaffilmark{3},
A. C. A. Boogert\altaffilmark{4},
G. M. Harper\altaffilmark{5},
D. T. Jaffe\altaffilmark{6},
K. R. Kulas\altaffilmark{7},
M. E. McKelvey\altaffilmark{8},
N. Ryde\altaffilmark{9},
W. Vacca\altaffilmark{4}
}

\altaffiltext{1}{Department of Physics and Astronomy, Johns Hopkins University, Baltimore, MD 21218, USA}
\altaffiltext{2}{Current address: Department of Astronomy, University of Michigan, Ann Arbor, MI 48109, USA}
\altaffiltext{3}{Department of Physics, University of California Davis, Davis, CA 95616, USA}
\altaffiltext{4}{USRA, SOFIA, NASA Ames Research Center, MS 232-11, Moffett Field, CA 94035, USA}
\altaffiltext{5}{School of Physics, Trinity College, Dublin 2, Ireland}
\altaffiltext{6}{Department of Astronomy, University of Texas, Austin, TX 78712, USA}
\altaffiltext{7}{Department of Physics, Santa Clara University, Santa Clara, CA 95053, USA}
\altaffiltext{8}{NASA Ames Research Center, Moffett Field, CA 94035, USA}
\altaffiltext{9}{Department of Astronomy and Theoretical Physics, Lund University, Lund, Sweden}

\begin{abstract}
We present high spectral resolution ($\sim$3~km~s$^{-1}$) observations of the $\nu_2$ ro-vibrational band of H$_2$O in the  6.086--6.135~$\mu$m range toward the massive protostar AFGL~2591 using the Echelon-Cross-Echelle Spectrograph (EXES) on the Stratospheric Observatory for Infrared Astronomy (SOFIA).  Ten absorption features are detected in total, with seven caused by transitions in the $\nu_2$ band of H$_2$O, two by transitions in the first vibrationally excited $\nu_2$ band of H$_2$O, and one by a transition in the $\nu_2$ band of H$_2^{18}$O.  Among the detected transitions is the $\nu_2$~1$_{1,1}$--0$_{0,0}$ line which probes the lowest lying rotational level of {\it para}-H$_2$O.  The stronger transitions appear to be optically thick, but reach maximum absorption at a depth of about 25\%, suggesting that the background source is only partially covered by the absorbing gas, or that the absorption arises within the 6~$\mu$m emitting photosphere.  Assuming a covering fraction of 25\%, the H$_2$O column density and rotational temperature that best fit the observed absorption lines are $N({\rm H_2O})=(1.3\pm0.3)\times10^{19}$~cm$^{-2}$ and $T=640\pm80$~K.
\end{abstract}

\section{Introduction} \label{sec_intro}

Water, despite being one of the most abundant species in the molecular interstellar medium (ISM), is difficult to observe in astrophysical objects due to its prevalence in the Earth's atmosphere \citep[see][for a comprehensive review of astronomical water observations]{vandishoeck2013}. Ground-based observations of H$_2$O have primarily targeted maser emission, most frequently the $J_{K_{a}K_{c}}=6_{1,6}$--5$_{2,3}$ transition near 22~GHz that was utilized in the initial detection of interstellar water \citep{cheung1969}, although some have also focused on rotational and ro-vibrational transitions out of high-lying rotational levels in the mid-IR \citep{pontoppidan2010} and near-IR \citep{najita2000,carr2004,salyk2008,indriolo2013_H2O}, respectively.  These latter observations are possible because many high-lying rotational levels are not significantly populated in the Earth's atmosphere, but their scope is limited to astrophysical sources with warm ($T>300$~K), dense gas.

Observations of the lowest-lying rotational levels of water---those able to probe cold gas---have required space-based observatories. The {\it Infrared Space Observatory}-Short Wavelength Spectrometer \citep[{\it ISO}-SWS;][]{kessler1996,degraauw1996} covered the $\nu_2$ ro-vibrational band (symmetric bending mode) of H$_2$O centered near 6~$\mu$m, and absorption out of the lowest-lying levels of the {\it ortho} and {\it para}  nuclear spin modifications (1$_{0,1}$ and 0$_{0,0}$, respectively) was detected toward several massive protostars \citep{boonman2003}.  Due to the low spectral resolution of the observing configuration though ($\lambda/\Delta\lambda\sim1400$ using SWS in AOT6 grating mode), these lines were significantly blended with absorption from other nearby H$_2$O lines, making the determination of level-specific column densities impossible.  Instead, the entire $\nu_2$ band was fit simultaneously assuming a single temperature to determine the total water column density, $N({\rm H_2O})$.  The {\it Submillimeter Wave Astronomy Satellite} \citep[SWAS;][]{melnick2000swas} provided much higher spectral resolution ($\lesssim1$~km~s$^{-1}$) and covered the 1$_{1,0}$--1$_{0,1}$ pure rotational transition of H$_2$O at 557~GHz.  This line was observed in both emission and absorption in multiple sources \citep[e.g.,][]{melnick2000,snell2000}, demonstrating the ability to probe cold water.  More recently, the study of low-lying rotational levels at high spectral resolution ($\sim0.5$~km~s$^{-1}$) has been facilitated by the Heterodyne Instrument for the Far-Infrared \citep[HIFI;][]{degraauw2010} on board the {\it Herschel Space Observatory} \citep{pilbratt2010}.  Water has been detected in both emission and absorption out of levels with $E\lesssim200$~K in several protostars \citep[e.g.,][]{vandishoeck2011,vandertak2013H2O}, and in absorption out of the 0$_{0,0}$ and 1$_{0,1}$ levels in the molecular ISM \citep[e.g.,][]{sonnentrucker2010,flagey2013}.  Observations that resolve the velocity structure of absorption lines are vital to both determining level-specific column densities, and understanding the dynamics of the absorbing/emitting regions.  This is especially important for protostars as such objects contain multiple dynamical components (e.g., disk, envelope, jets, outflows, shocks).

\begin{figure*}
\epsscale{1.1}
\plotone{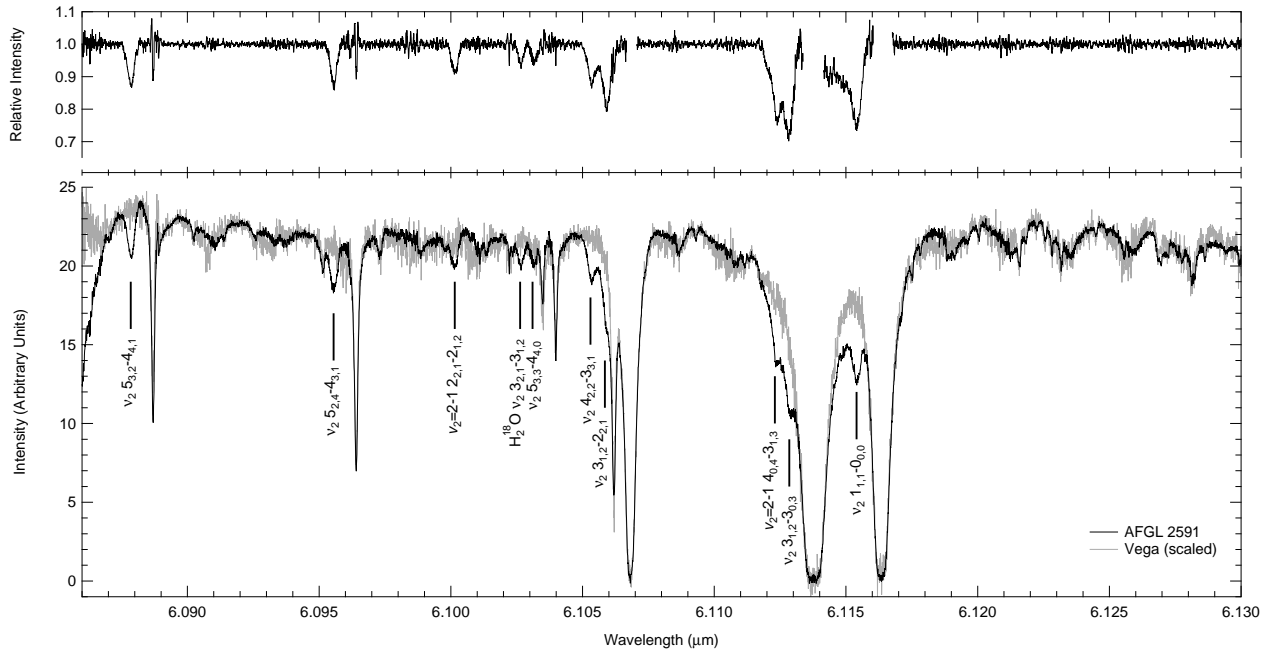}
\caption{{\bf Bottom:} Spectra of AFGL 2591 (black) and Vega (gray) scaled to match in arbitrary intensity units.  Water absorption features in AFGL~2591 are marked by vertical lines and labeled accordingly.  They are blue-shifted by about 40~km~s$^{-1}$ with respect to their telluric counterparts due to the Earth's motion and gas velocity. {\bf Top:} Spectrum of AFGL~2591 following division by boxcar averaged Vega spectrum and normalization. Gaps correspond to regions where the atmosphere is opaque, and poor removal of moderately strong atmospheric lines can be seen as spikes to the right of the astrophysical lines (e.g., near 6.089~$\mu$m.}
\label{fig_fullspectrum}
\end{figure*}

AFGL~2591 is a region of ongoing high-mass star formation.  The radio continuum source VLA~3 \citep[for a description of sources see][and references therein]{torrelles2014} is the brightest mid-IR source in the region and drives a bipolar outflow. It was previously observed at 5--7~$\mu$m as part of the aforementioned {\it ISO}-SWS study, from which \citet{boonman2003} reported a best-fit water column density of $N({\rm H_2O})=(3.5\pm1.5)\times10^{18}$~cm$^{-2}$ at $T=450_{-150}^{+250}$~K, assuming a Doppler line width of 5~km~s$^{-1}$.  Known kinematic components associated with AFGL~2591 include: (1) the protostellar envelope with systemic velocity $-5.5$~km~s$^{-1}$ in the local standard of rest (LSR) frame \citep{vandertak1999}; (2) a blue-shifted outflow at $-25$~km~s$^{-1}\lesssim v_{\rm LSR}\lesssim -6$~km~s$^{-1}$ \citep{emprechtinger2012,vandertak2013H2O}; (3) a red-shifted outflow at $3$~km~s$^{-1}\lesssim v_{\rm LSR}\lesssim 15$~km~s$^{-1}$ \citep{lada1984}; (4) material entrained in a jet or wind, indicated by absorption in the $v=1$--0 band of $^{12}$CO that extends to $-196$~km~s$^{-1}$ in the form of an asymmetric blue wing \citep{vandertak1999}. There is also evidence for a rotating disk around VLA~3, although gas velocities in the disk match those of the envelope \citep{vandertak2006H2O,wang2012}.  The interested reader is referred to cartoon pictures \citep{vandertak1999,vandertak2006H2O,vanderwiel2011,sanna2012,wang2012} and multi-wavelength images \citep{johnston2013} of AFGL~2591 to gain a better understanding of the region.  Because of the low spectral resolution of the {\it ISO}-SWS observations, it has been impossible to say with certainty which component gives rise to the H$_2$O absorption.  To do so---and to better constrain the water column density and rotational temperature---we have targeted multiple transitions in the $\nu_2$ band of H$_2$O using the Echelon-Cross-Echelle Spectrograph \citep[EXES;][]{richter2010} on board the Stratospheric Observatory for Infrared Astronomy \citep[SOFIA;][]{young2012}.  

SOFIA operates at altitudes above 39,000~ft (11887~m), where the precipitable water vapor overburden is routinely less than 0.02~mm.  Under these conditions the 6~$\mu$m region of the Earth's atmosphere is no longer opaque, as it is from the ground. EXES provides high spectral resolution ($\sim3$~km~s$^{-1}$) capabilities in the 4.5--28.3~$\mu$m range, making it well-suited for velocity-resolved observations of individual ro-vibrational transitions of the $\nu_2$ band of H$_2$O in astrophysical sources.  We present here the first spectrally resolved detections of 10 absorption lines from transitions in the $\nu_2$ bands of H$_2$O and H$_2^{18}$O, including a detection probing the ground {\it para} level, 0$_{0,0}$.  This highlights the opportunity to further probe water in cold molecular clouds without the need for a space-based observatory by utilizing EXES on SOFIA.

\section{Observations and Data Reduction}

AFGL~2591 VLA 3 was observed using EXES on board SOFIA at an altitude of 43000~ft (13106~m) on 2014~Apr~10 (UT) as part of instrument commissioning observations. Spectra were acquired in cross-dispersed high-resolution mode with a central wavelength of 6.1125~$\mu$m, using a slit length of 9\farcs9, and a slit width of 1\farcs9 to provide a resolving power (resolution) of 86,000 (3.5~km~s$^{-1}$), with the resolution element sampled by 8 pixels. The telescope was nodded after every 27 seconds of integration, enabling subtraction of telluric emission lines, and the total exposure time for AFGL~2591 was 1134~s.  Prior to observing the target, a calibration sequence was taken using the same wavelength setting and slit width. This sequence consisted of observations of the internal blackbody unit set to 260$\pm$0.1~K and of blank sky \citep{lacy2002}. These calibration frames were used to correct for blaze efficiency, pixel-to-pixel sensitivity variations, and also to provide a first order flux calibration using the expected photon counts from the blackbody unit.  The bright star Vega was observed during a flight leg 2~hr before the science observations at the same altitude and air mass as AFGL~2591 for use as a telluric standard star.  The observing sequence was similar to that employed for AFGL~2591.

Data were processed using the {\tt Redux} pipeline \citep{clarke2014} with the {\tt fspextool} software package---a modification of the Spextool package \citep{cushing2004}---which performs source profile construction, extraction and background aperture definition, optimal extraction, and wavelength calibration for EXES data. The preliminary wavelength scale output from the pipeline was refined by shifting the wavelength solution so that telluric water absorption lines in the AFGL~2591 continuum best matched entries in the HITRAN database \citep{hitran2012}. This wavelength calibration is accurate to within $\pm0.3$~km~s$^{-1}$. Spectra extracted from individual orders of the echellogram were combined to form a single continuous spectrum in the wavelength range 6.085--6.135~$\mu$m for both AFGL~2591 and Vega as shown in the bottom panel of Figure \ref{fig_fullspectrum}.  To remove atmospheric features, the spectrum of AFGL~2591 was divided by a smoothed (11 point boxcar average) Vega spectrum. The resulting ratioed spectrum was then divided by a boxcar average to the continuum level extrapolated across absorption lines to produce the normalized spectrum in the top panel of Figure \ref{fig_fullspectrum}. Figure \ref{fig_transitions} shows the spectra for each transition shifted into the LSR frame following conversion from wavelength to line-of-sight velocity.

\section{Analysis}
\begin{figure}
\epsscale{1.25}
\plotone{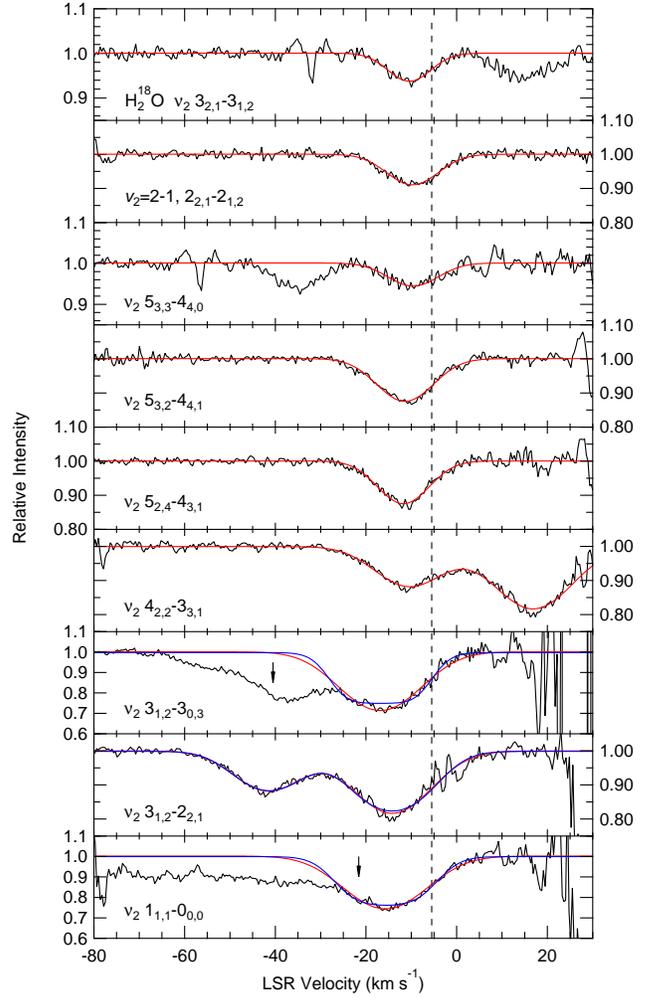}
\caption{Individual H$_2$O absorption lines as functions of LSR velocity, arranged in order of increasing lower-state energy (except the H$_2^{18}$O line).  The vertical dashed line shows the systemic velocity of AFGL~2591 determined from various molecular emission lines \citep[$v_{\rm LSR}=-5.5$~km~s$^{-1}$;][]{vandertak1999}.  Red and blue curves show fits with $f_c=1$ and $f_c=0.25$, respectively.  For optically thin lines both fits are nearly identical, so only the $f_c=1$ fits are shown. At $v_{\rm LSR}\lesssim-25$~km~s$^{-1}$ in the $\nu_2$ 1$_{1,1}$--0$_{0,0}$ spectrum we consider the lower continuum level to be caused by poor removal of strong atmospheric lines.  The arrow at $-$22~km~s$^{-1}$ marks the expected position of the $v_2=2$--1~3$_{1,3}$--2$_{0,2}$ transition given a line center of $-$11~km~s$^{-1}$.  In the 
$\nu_2$~3$_{1,2}$--3$_{0,3}$ spectrum the arrow at $-$41~km~s$^{-1}$ marks the expected position of the $v_2=2$--1~4$_{0,4}$--3$_{1,3}$ transition, and we consider the feature from about $-32$~km~s$^{-1}$ to $-47$~km~s$^{-1}$ to be a detection of this line, although it is also affected by a sloping baseline due to poor removal of atmospheric lines.}
\label{fig_transitions}
\end{figure}

Absorption lines in Figure \ref{fig_transitions} were fit using a function of the form
\begin{equation}
I=I_0\left[1-f_c\left[1-\exp\left(-\tau_0\exp\left(-\frac{(v-v_{\rm LSR})^2}{2\sigma_{v}^2}\right)\right)\right]\right],
\label{eq_fitting}
\end{equation}
where $I_0$ is the continuum level, $f_c$ is the fraction of the background source covered by absorbing material, and the optical depth is assumed to have a Gaussian profile with optical depth at line center $\tau_0$, LSR velocity at line center $v_{\rm LSR}$, and velocity dispersion $\sigma_v$.  The $\nu_2$~3$_{1,2}$--2$_{2,1}$ and 
$\nu_2$~4$_{2,2}$--3$_{3,1}$ transitions were fit simultaneously (second and fourth panels from the bottom in Figure \ref{fig_transitions}) using a version of this function modified for multiple lines.  The $\nu_2$~5$_{3,3}$--4$_{4,0}$ and H$_2^{18}$O $\nu_2$~3$_{2,1}$--3$_{1,2}$ transitions (third-from-top and top panels) were fit separately despite their proximity to each other as there is no significant blending of the features. Due to interfering features near the $\nu_2$~1$_{1,1}$--0$_{0,0}$ and $\nu_2$~3$_{1,2}$--3$_{0,3}$ transitions (bottom and third-from-bottom panels; see figure caption for explanation of features) only data at $v_{\rm LSR}\geq-23$~km~s$^{-1}$ were used to constrain fits of these transitions.  As the absorption features are broad with respect to the instrumental spectral resolution, we expect the lines to be resolved and convert the optical depth profile fit to a column density profile via
\begin{equation}
dN/dv=\tau(v)\frac{g_l}{g_u}\frac{8\pi}{A\lambda^3},
\label{eq_tautoN}
\end{equation}
under the assumption that the absorption is unsaturated, where $g_l$ and $g_u$ are statistical weights in the lower and upper states respectively, $A$ is the spontaneous emission coefficient, and $\lambda$ is the transition wavelength.  We then integrate over the absorption feature in velocity space to determine the column density in the lower state of the observed transition.

\begin{deluxetable*}{rcccccccc}
\tabletypesize{\scriptsize}
\tablecaption{Observed Water Transitions and Inferred Parameters\label{tbl_transitions}}
\tablehead{\colhead{Transition\tablenotemark{a}} & \colhead{Wavelength} & \colhead{$E_l/k_b$} & \colhead{$g_l$} & \colhead{$A$} & \colhead{$v_{\rm LSR}$} &  \colhead{$\sigma_v$} & \colhead{$\tau_0$} & \colhead{$\ln(f_cN_{l}/g_{l})$} \\
 & \colhead{($\mu$m)} & \colhead{(K)} & & \colhead{(s$^{-1}$)} & \colhead{(km~s$^{-1}$)} & \colhead{(km~s$^{-1}$)} & & 
}
\startdata
 & & & & & \multicolumn{3}{c}{Fit Results for $f_c=1$} & \\
1--0~5$_{3,2}$--4$_{4,1}$ & 6.0887005 & 702.3 & 27 & 0.35 & -11.4 & 6.0 & 0.13$\pm0.03$ & 35.2$\pm$0.3  \\
1--0~5$_{2,4}$--4$_{3,1}$ & 6.0964081 & 552.3 & 9 & 0.73 & -11.7 & 5.5 & 0.13$\pm0.03$ & 35.5$\pm$0.3 \\
2--1~2$_{2,1}$--2$_{1,2}$ & 6.1009690 & 2412.9 & 15 & 5.92 & -9.7 & 5.5 & 0.09$\pm0.03$ & 32.7$\pm$0.4  \\
H$_2^{18}$O~1--0~3$_{2,1}$--3$_{1,2}$ & 6.1034870 & 248.7 & 21 & 6.40 & -10.4 & 4.7 & 0.06$\pm0.03$ & 31.8$\pm$0.5 \\
1--0~5$_{3,3}$--4$_{4,0}$ & 6.1039868 & 702.3 & 9 & 0.34 & -9.6 & 5.5 & 0.06$\pm0.03$ & 35.4$\pm$0.6  \\
1--0~4$_{2,2}$--3$_{3,1}$ & 6.1061925 & 410.4 & 7 & 0.53 & -10.5 & 7.3 & 0.12$\pm0.03$ & 36.2$\pm$0.3  \\
1--0~3$_{1,2}$--2$_{2,1}$ & 6.1068262 & 194.1 & 15 & 1.12 & -14.2 & 8.5 & 0.20$\pm0.04$ & 35.2$\pm$0.2  \\
2--1~4$_{0,4}$--3$_{1,3}$ & 6.1131638 & 2502.7 & 7 & 15.8 & -8.4 & 4.4 & 0.12$\pm0.03$ & 32.2$\pm$0.3 \\
1--0~3$_{1,2}$--3$_{0,3}$ & 6.1137707 & 196.8 & 21 & 6.24 & -16.9 & 8.5 & 0.34$\pm0.04$ & 34.0$\pm$0.1 \\
1--0~1$_{1,1}$--0$_{0,0}$ & 6.1163311 & 0 & 1 & 7.46 & -15.7 & 8.7 & 0.30$\pm0.04$ & 35.7$\pm$0.1  \\
\hline
 & & & & & \multicolumn{3}{c}{Fit Results for $f_c=0.25$} & \\
1--0~5$_{3,2}$--4$_{4,1}$ & 6.0887005 & 702.3 & 27 & 0.35 & -11.4 & 5.6 & 0.68$^{+0.27}_{-0.21}$ & 35.4$^{+0.4}_{-0.3}$   \\
1--0~5$_{2,4}$--4$_{3,1}$ & 6.0964081 & 552.3 & 9 & 0.73 & -11.8 & 5.2 & 0.68$^{+0.27}_{-0.21}$ & 35.6$^{+0.4}_{-0.3}$   \\
2--1~2$_{2,1}$--2$_{1,2}$ & 6.1009690 & 2412.9 & 15 & 5.92 & -9.8 & 5.3 & 0.44$^{+0.21}_{-0.18}$ & 32.8$^{+0.5}_{-0.4}$   \\
H$_2^{18}$O~1--0~3$_{2,1}$--3$_{1,2}$ & 6.1034870 & 248.7 & 21 & 6.40 & -10.4 & 4.5 & 0.29$^{+0.17}_{-0.15}$ & 31.8$^{+0.6}_{-0.5}$   \\
1--0~5$_{3,3}$--4$_{4,0}$ & 6.1039868 & 702.3 & 9 & 0.34 & -9.6 & 5.3 & 0.25$^{+0.17}_{-0.14}$ & 35.4$^{+0.7}_{-0.6}$   \\
1--0~4$_{2,2}$--3$_{3,1}$ & 6.1061925 & 410.4 & 7 & 0.53 & -10.3 & 6.9 & 0.62$^{+0.25}_{-0.20}$ & 36.4$^{+0.4}_{-0.3}$   \\
1--0~3$_{1,2}$--2$_{2,1}$ & 6.1068262 & 194.1 & 15 & 1.12 & -14.2 & 7.3 & 1.20$^{+0.51}_{-0.34}$ & 35.5$^{+0.4}_{-0.3}$   \\
2--1~4$_{0,4}$--3$_{1,3}$ & 6.1131638 & 2502.7 & 7 & 15.8 & -8.4 & 4.1 & 0.57$^{+0.24}_{-0.19}$ & 32.4$^{+0.4}_{-0.3}$   \\
1--0~3$_{1,2}$--3$_{0,3}$ & 6.1137707 & 196.8 & 21 & 6.24 & -16.8 & 5.7 & $>5.1$ & $>35.0$   \\
1--0~1$_{1,1}$--0$_{0,0}$ & 6.1163311 & 0 & 1 & 7.46 & -15.6 & 6.5 & $>3.0$ & $>36.3$
\enddata
\tablecomments{Columns 2, 3, 4, and 5 give the transition wavelength, $\lambda$, lower state energy, $E_l/k_b$, lower-state statistical weight, $g_l$, and the spontaneous emission coefficient, $A$, respectively. Columns 6, 7, and 8 give the best-fit parameters from equation (\ref{eq_fitting}), with covering fractions of 100\% and 25\% assumed in the top and bottom halves of the table, respectively. Uncertainties in $v_{\rm LSR}$ and $\sigma_v$ are estimated to be 0.3~km~s$^{-1}$ at the 1$\sigma$ level.  Uncertainties in $\tau_0$ assume that the uncertainty in the absorption depth at line center is equal to the root mean square noise level of 0.03 in the continuum of the ratioed spectrum prior to the removal of baseline fluctuations. 
The Doppler parameter, $b$, and line full width at half maximum, FWHM, are related to $\sigma_v$ via the equations $b=\sigma_v\sqrt{2}$ and ${\rm FWHM}=\sigma_{v}2\sqrt{2\ln(2)}$.}
\tablenotetext{a}{Transition labels are given as $v_{2}'$--$v_{2}''$~$J_{K_{a}'K_{c}'}'$--$J_{K_{a}''K_{c}''}''$, where a single prime denotes the upper state and double prime denotes the lower state, and both the $v_1$ and $v_3$ vibrational quantum numbers are omitted as they are 0 in all cases.}
\end{deluxetable*}
\normalsize

\begin{figure*}
\epsscale{1.2}
\plotone{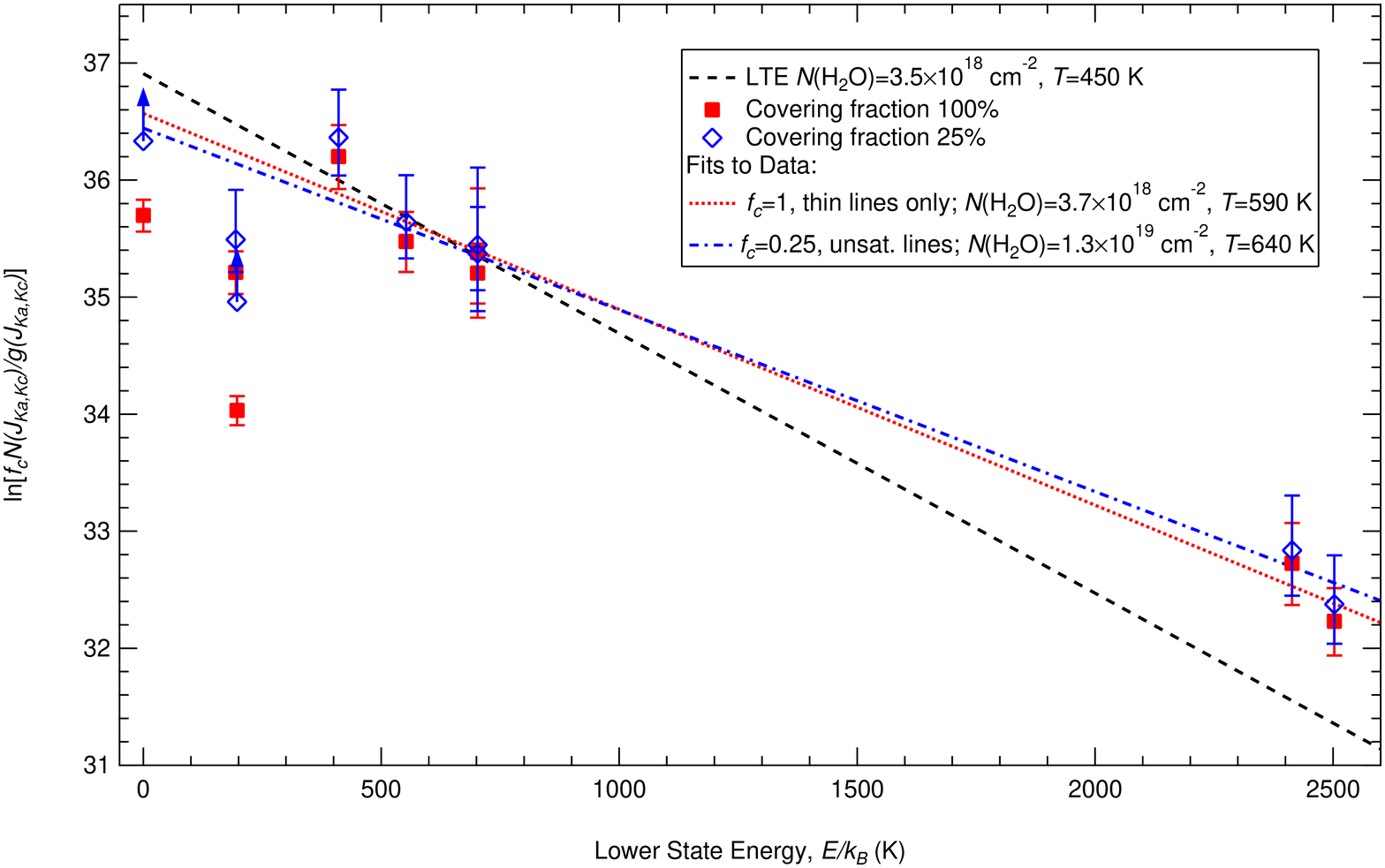}
\caption{Rotation diagram for the 9 detected H$_2$O transitions.  The black dashed line is for $N({\rm H_{2}O})=3.5\times10^{18}$~cm$^{-2}$ and LTE at 450~K as reported by \citet{boonman2003}.  Red squares and open blue diamonds denote covering fractions of 1 and 0.25, respectively. Lower limits are indicated by upward-pointing arrows. Note that the $y$-axis shows $\ln(f_{c}N(J_{K_{a}K_{c}})/g_{K_{a}K_{c}})$, (i.e., is scaled to the average column density in front of the entire background source). For optically thin absorption lines the value does not change significantly with covering fraction, but for optically thick lines (e.g., $\nu_2$~1$_{1,1}$--0$_{0,0}$) the value increases substantially as $f_c$ approaches the absorption depth. The red dotted line marks $T=590$~K and $N({\rm H_{2}O})=3.7\times10^{18}$~cm$^{-2}$ for $f_c=1$, and is the best fit to levels with $E/k_B>300$~K. The blue dash-dot line marks $T=640$~K and $N({\rm H_{2}O})=1.3\times10^{19}$~cm$^{-2}$ for $f_c=0.25$, and is the best fit to the 7 unsaturated transitions.} 
\label{fig_rotdiag}
\end{figure*}

Initial fits were made assuming that the absorbing gas completely covers the background source ($f_c=1$), and are shown as red curves in Figure \ref{fig_transitions}.  Resulting parameters for each transition are reported in Table \ref{tbl_transitions}.  Inferred column densities are converted to $\ln(f_{c}N/g_l)$ and plotted versus lower state energy as red squares in Figure \ref{fig_rotdiag}. The black dashed line shows the relationship expected for $N({\rm H_2O})=3.5\times10^{18}$~cm$^{-2}$ in local thermodynamic equilibrium (LTE) at $T=450$~K \citep{boonman2003}.  Column densities in the 3$_{3,1}$, 4$_{3,1}$, 4$_{4,1}$, and 4$_{4,0}$ states are in agreement with predictions based on these values, but the states with $E/k_{B}<300$~K are below predicted values, while vibrationally excited states are above predicted values. No single temperature provides a good fit to all nine points, but if we exclude levels with $E/k_{B}<300$~K then we find $N({\rm H_2O})=(3.7\pm0.8)\times10^{18}$~cm$^{-2}$ in LTE at $T=590\pm50$~K (marked by the red dotted line in Figure \ref{fig_rotdiag}).

Assuming $T$ and $N({\rm H_2O})$ from \citet{boonman2003} the $\nu_2$~3$_{1,2}$--3$_{0,3}$ transition should be saturated at line center, which is not the case.  If the background source is partially covered by absorbing material, or if the water absorption arises within the 6~$\mu$m emitting photosphere, then it is possible for optically thick, saturated lines to cause only a fractional decrease in the continuum level.  The minimum possible fractional coverage is equal to the depth of the strongest absorption line under consideration, in our case the $\nu_2$~3$_{1,2}$--3$_{0,3}$ transition.  However, the perceived astrophysical absorption from this transition is greatly affected by its telluric counterpart, and the line depth is highly dependent on the atmospheric division procedure.  The same is true for the next strongest line, the $\nu_2$~1$_{1,1}$--0$_{0,0}$ transition.  We use both lines in estimating the minimum covering fraction, choosing $f_c=0.25$, but caution that this is highly uncertain. Fits using $f_c=0.25$ are shown as blue curves in Figure \ref{fig_transitions}, and resulting parameters are again in Table \ref{tbl_transitions}.  Inferred column densities are plotted as open blue diamonds in Figure \ref{fig_rotdiag}.  Scaling by the covering fraction demonstrates that column densities inferred from optically thick transitions increase much more than those inferred from optically thin transitions as $f_c$ decreases, but at $f_c=0.25$ the saturated $\nu_2$~3$_{1,2}$--3$_{0,3}$ and  $\nu_2$~1$_{1,1}$--0$_{0,0}$ transitions only provide lower limits on column densities. A fit to the 7 unsaturated transitions is shown by the blue dash-dot line in Figure \ref{fig_rotdiag}, and corresponds to $N({\rm H_2O})=(1.3\pm0.3)\times10^{19}$~cm$^{-2}$ in LTE at $T=640\pm80$~K.

\section{Discussion}

The pure rotational 1$_{1,1}$--0$_{0,0}$ transition of H$_2$O at 1113.343~GHz (269.3~$\mu$m) is seen in both emission and absorption toward AFGL~2591 \citep{vandertak2013H2O,kazmierczak-barthel2014,choi2015}.  The emission component arises in the protostellar envelope, and the absorption components in the blue-shifted outflow and foreground gas.  This absorption probes the same quantum state as the $\nu_2$~1$_{1,1}$--0$_{0,0}$ transition, but a direct comparison is hindered by several effects.  The emitting regions at 6~$\mu$m and 1.1~THz ($\sim$270~$\mu$m) are likely different, so that gas probed by one transition may not necessarily be probed by the other.  Our EXES observations used a 9\farcs9 by 1\farcs9 slit centered on VLA~3, while HIFI observations at 1113~GHz have a roughly circular beam with FWHM~$\sim19$\arcsec, again meaning that different regions are being probed.  Finally, emission from the ($v_{1}v_{2}v_{3})$~$J_{K_a,K_c}$=(000)~1$_{1,1}$ state (53.4~K above ground) is extremely strong and interferes with the absorption features at 1113~GHz, whereas emission from the (010)~1$_{1,1}$ state (2352~K above ground) is not observed, leaving the 6.1163311~$\mu$m line unobscured.  All of these effects must be considered when comparing any results from HIFI and EXES.

As described in Section \ref{sec_intro}, different components of AFGL~2591 are distinguished by line-of-sight velocities, line profiles, and observed molecules. The H$_2$O absorption lines are centered at roughly $-16$~km~s$^{-1}$ with FWHM~$\sim20$~km~s$^{-1}$ for the three lowest-lying levels and shift to  $-11$~km~s$^{-1}$ with FWHM~$\sim13$~km~s$^{-1}$ for the higher-lying levels, best matching the blue-shifted outflow. These line profiles are most similar to absorption seen in HCN and C$_2$H$_2$ \citep[][observed at 13~$\mu$m]{knez2006} as well as specific absorption components of $^{13}$CO and vibrationally excited $^{12}$CO \citep[][observed at 4.7~$\mu$m]{mitchell1989,vandertak1999}, all of which are thought to arise in hot, dense gas close to the central protostar. Several 22~GHz H$_2$O masers are observed throughout the region, with many concentrated in the walls of the blue-shifted outflow associated with VLA~3 \citep{trinidad2003,sanna2012,torrelles2014}.  Although maser velocities are primarily at $v_{\rm LSR}\leq-18$~km~s$^{-1}$, absorption by gas in small, shocked clumps could explain the small covering fraction required by our analysis.  Given the above, we posit that the H$_2$O absorption observed with EXES arises in hot, dense gas at the base of the blue-shifted outflow.

Water emission attributed to the outflow component is observed with HIFI in multiple transitions as a broad feature \citep{vandertak2013H2O,kazmierczak-barthel2014,choi2015}, but is centered closer to the systemic velocity of the envelope than the absorption we see.  Analysis of the H$_2$O outflow emission by \citet{choi2015} indicates a temperature of $T\sim70$--90~K and column density of $N({\rm H_2O})\sim4\times10^{13}$~cm$^{-2}$, similar to results found by \citet{karska2014} who analyzed unresolved H$_2$O absorption observed with PACS on {\it Herschel}, finding $T=160\pm130$~K and $N({\rm H_2O})\sim10^{14}$~cm$^{-2}$.  These values are significantly below what we find, suggesting that EXES and HIFI/PACS observations trace different components. Between the broad and narrow water emission observed with HIFI, the broad absorption observed with EXES, and the abundant maser spots, it is evident that the water-containing gas in AFGL~2591 is both spatially and kinematically complex.  Interpreting the various observations of H$_2$O in AFGL~2591 in unison will require utilizing a physical model that includes radiative transfer accounting for absorption, stimulated and spontaneous emission, and collisional (de-)excitation, as well as kinematic and geometric effects.

\section{Summary}
We have detected ten absorption features arising from warm gas in AFGL~2591 caused by ro-vibrational transitions of water, including seven from the $\nu_2$ band of H$_2$O, two from the vibrationally excited $\nu_2$ band of H$_2$O, and one from the $\nu_2$ band of H$_2^{18}$O.  Among the detected absorption lines is the $\nu_2$~1$_{1,1}$--0$_{0,0}$ transition at 6.1163311~$\mu$m, which probes the ground state of {\it para}-H$_2$O. Relative strengths of absorption features are suggestive of a covering fraction less than 1 (or absorption arising within the 6~$\mu$m emitting photosphere), with a limit of $f_c\geq0.25$ set by the depth of the strongest absorption features.  Analysis of the level populations assuming $f_c=0.25$ results in $N({\rm H_2O})=(1.3\pm0.3)\times10^{19}$~cm$^{-2}$ for LTE at $T=640\pm80$~K.  Line profiles best match the blue-shifted outflow component, and we ascribe the absorption to hot, dense gas at the base of the outflow. The temperature and column density inferred by our analysis are much larger than those reported by \citet{choi2015} for the outflow component observed in H$_2$O emission by HIFI.  Uncertainty in whether the EXES and HIFI observations are probing the same gas makes the combined interpretation of both datasets difficult, and a physical model of AFGL~2591 that includes radiative transfer will be necessary for such an analysis.  Clearly though, observations of the $\nu_2$ ro-vibrational band of H$_2$O at high spectral resolution---observations uniquely achievable with EXES on SOFIA---add important information for interpreting this region and others.

MJR and CND acknowledge Collaborative Agreement NNX13AI85A between UCD and NASA Ames for its support and support of EXES development. Many thanks to the anonymous referee.

%%%%%%%%%%%%%%%%%%%%%%%%%%%%%%Tables%%%%%%%%%%%%%%%%%%%%%%%%%%%%%%%%%%%%%%%%%%%

%%%%%%%%%%%%%%%%%%%%%%%%%%%%%%Figures%%%%%%%%%%%%%%%%%%%%%%%%%%%%%%%%%%%%%%%%%%
\end{document}